\RequirePackage{fix-cm}
\documentclass[smallextended]{svjour3}       
\smartqed  
\usepackage{graphicx}
\usepackage{amssymb}
\usepackage[draft]{hyperref}

\usepackage[dvipsnames]{xcolor} 

\newcommand{\lam}{$\lambda$}
\newcommand{\1}{\footnotesize I\normalsize}
\newcommand{\2}{\footnotesize II\normalsize}
\newcommand{\3}{\footnotesize III\normalsize}
\newcommand{\4}{\footnotesize IV\normalsize}
\newcommand{\5}{\footnotesize V\normalsize}

\newcommand{\kms}{km\,s$^{-1}$}
\newcommand{\po}{$\phantom{o}$}
\begin{document}

\title{A near-UV reconnaissance of metal-poor massive stars
}


\author{Chris Evans         \and
        Wagner Marcolino    \and 
        Jean-Claude Bouret \and
        Miriam Garcia
}

\authorrunning{Evans et al.} 

\institute{C. Evans\at
European Space Agency (ESA), ESA Office, Space Telescope Science Institute, 3700 San Martin Drive, Baltimore, MD 21218, USA \\
              \email{chevans@stsci.edu}           
           \and
           W. Marcolino \at
Universidade Federal do Rio de Janeiro, Observat\'orio do Valongo, 
Ladeira Pedro Ant\^{o}nio, 43, CEP 20.080-090, Rio de Janeiro, Brazil
           \and
           J.-C. Bouret \at
           Aix-Marseille Univ, CNRS, CNES, LAM, Marseille, France
           \and
           M. Garcia \at
           Centro de Astrobiolog\'{i}a, CSIC-INTA, Crtra. de Torrej\'{o}n a Ajalvir km 4, E-28850, Torrej\'{o}n de Ardoz (Madrid), Spain
}

\date{Accepted: 8 August 2023}

\maketitle

\begin{abstract}
  We use synthetic model spectra to investigate the potential of
  near-ultraviolet (3000-4050\,\AA) observations of massive O-type
  stars. We highlight the He~{\scriptsize I} \lam3188 and
  He~{\scriptsize II} \lam3203 pair as a potential temperature
  diagnostic in this range, supported by estimates of gravity using
  the high Balmer series lines. The near-ultraviolet also contains
  important metallic lines for determinations of chemical abundances
  (oxygen in particular) and estimates of projected rotational
  velocities for O-type spectra. Using the model spectra we present
  performance estimates for observations of extragalactic massive
  stars with the Cassegrain U-Band Efficient Spectrograph (CUBES) now
  in construction for the Very Large Telescope. The high efficiency of
  CUBES will open-up exciting new possibilities in the study of
  massive stars in external galaxies. For instance, CUBES will provide new
  insights into the physical properties of O-type stars, including
  oxygen abundances, in metal-poor irregular galaxies at $\sim$1\,Mpc
  from integrations of just 2-3\,hrs. Moreover, CUBES will bring
  quantitative spectroscopy of more distant targets within reach for
  the first time, such as the O-type star ($V$\,$\sim$\,21.5\,mag) in
  Leo~P (at 1.6\,Mpc) in only half a night of observations.

\end{abstract}

\section{Introduction}
\label{intro}

Ground-based spectroscopy of OB-type stars has traditionally been obtained
in the \lam3950--4750\AA\ range for spectral classification (e.g.
\cite{wf00}) and estimates of physical parameters (temperatures,
gravities), and of the H$\alpha$ line to investigate their stellar
winds (e.g. \cite{puls96}). In contrast, the shorter wavelengths
accessible from the ground, down to the atmospheric cut-off, have
received significantly less attention, partly due to the challenges of
the reduced atmospheric transmission combined with the limited
efficiency of available instrumentation.

Development of the Cassegrain U-Band Efficient Spectrograph
(CUBES) instrument for the Very Large Telescope (VLT) provided us with
the motivation to investigate the potential of observations of massive
stars in the near ultraviolet (UV).  The CUBES design offers a
potential tenfold gain in end-to-end efficiency at
\lam\,$<$\,3400\,\AA\ (incl. the telescope and atmosphere)
compared to the existing Ultraviolet and Visible Echelle Spectrograph
(UVES). In brief, the CUBES design provides a spectral resolving power
of $R$\,$\ge$\,20,000 over the 3000-4050\,\AA\ range, with provision
of a second, lower-resolution option with $R$\,$\sim$\,7,000 \cite{zanutta22}.

Here we investigate the potential performance of CUBES for studies of
massive stars. In Sect.~2 we briefly review past studies of massive
stars shortwards of 4000\,\AA.  Motivated by the presence of a
relatively strong He~\2 line (\lam3203) and a range of He~\1 lines, in
Sect.~3 we use synthetic model spectra to qualitatively investigate
the sensitivity of near-UV lines to the physical parameters of massive
stars. In Sect.~4 we focus on the possibility of estimating oxygen
abundances in massive stars from near-UV observations. In Sect.~5 we
investigate the potential performance of CUBES to study massive stars
in Local Group galaxies and beyond, with concluding remarks given in
Sect.~6.

\section{Near-ultraviolet spectroscopy of massive stars}

An early ground-based study of massive stars at
\lam\,$<$\,4000\,\AA\ used photographic observations of $\epsilon$~Ori
(B0 Ia), which included a detailed linelist that extended as far
bluewards as 3550\,\AA\ \cite{lamers72}.  A first quantitative
investigation of the He~\1 \lam3188 and He~\2 \lam3203 lines employed
photographic observations from the 2.2-m telescope on Mauna Kea of 19
O- and B0.5-type stars \cite{m75}. This included equivalent
width-measurements of both lines in the sample of stars, and
comparisons with the predictions of non-LTE model atmospheres for
He~\1 \lam3889 and He~\2 \lam3203, finding generally good agreement
for the trend of the He~\1 lines and the values for the He~\2 line
(except for the two hottest stars). Given the strong response to
temperature of both lines, this was a first indication of the
potential of these two lines to be used in tandem as a temperature
diagnostic. The only other example known to us of observations in
this region prior to the use of digital detectors is observations of
the He~\1 \lam3188 line in $\sim$30 early-type stars with the {\em
  Copernicus} satellite \cite{dufton}.

Digital detectors transformed observational astronomy, but their
performance in the near UV has still been a limiting factor compared to longer
wavelengths, and so there remain relatively few studies of massive
stars in this region.  Observations covering 3250-4750\,\AA\ of
massive O-type and Wolf--Rayet stars in NGC\,3603 with the Faint
Object Spectrograph on the {\em Hubble Space Telescope (HST)} revealed
some of the key features in the near UV \cite{drissen95}. Shortwards
of 4000\,\AA, the O-type spectra are dominated by the high Balmer
series until the Balmer limit, with a He~\1 line at \lam3820. Going to
even shorter wavelengths, in the hottest (O3-type) stars in the {\em
  HST} observations, absorption from O~\4 \lam\lam3381-85, 3412 and
N~\4 \lam\lam3479-83-85 are also seen, with the N~\4 blend displaying
a strong P~Cygni profile in the hydrogen-rich WN-type spectra.
Examples of some of the weak He~\1 lines present in the
3800-4000\,\AA\ range can also be seen in slightly higher-resolution
($\sim$2\,\AA) spectroscopy of four late O-type supergiants (from
\cite{wh00}).

\begin{figure}
  \includegraphics[width=12cm]{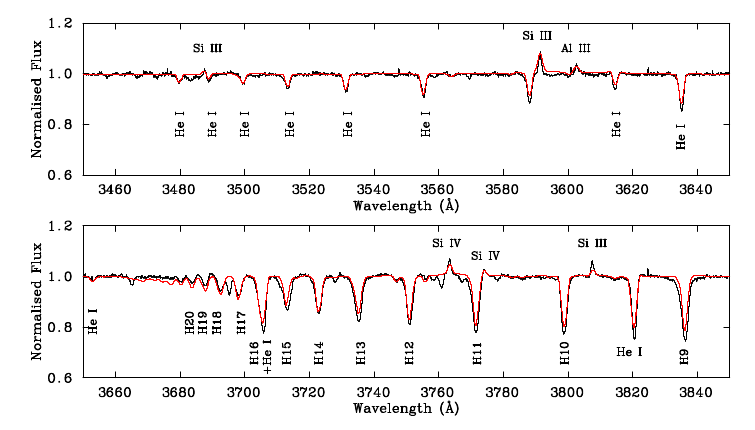}
  \caption{{\it Black spectrum:} UVES spectrum of HDE\,269896, an ON9.7~Ia+ star in the Large
    Magellanic Cloud.  The data have been smoothed and rebinned to
    match the CUBES high-resolution mode (with $R$\,$\sim$22,000).
    Identified lines in the upper panel are: Al~{\scriptsize III}
    \lam3602; He~{\scriptsize I} \lam\lam3479, 3488, 3499, 3513, 3531,
    3554, 3614, 3634; Si~{\scriptsize III} \lam\lam3487, 3590.  In
    addition to the high Balmer lines (H9-20), identified lines in the
    lower panel are: He~{\scriptsize I} \lam\lam3652, 3820;
    Si~{\scriptsize III} \lam3807; Si~{\scriptsize IV} \lam\lam3762, 3773.
  {\it Red spectrum:} Adopted {\sc cmfgen} model from a combined UV and optical analysis of the UVES data (taken fron \cite{ecfh04}, see text for details).}
\label{hd269896}
\end{figure}

To illustrate some of the spectral lines present in this range, in
Fig.~\ref{hd269896} we show the far-blue UVES spectrum of HDE\,269896
(taken from \cite{ecfh04}), smoothed and rebinned to the high-resolution
mode of CUBES ($\Delta$\lam\,$=$\,0.14\,\AA, sampled by 2.3 pixels, so
0.06\,\AA/pixel). 

The region below $\sim$3400\,\AA\ in the UVES data was of limited use
because of low signal-to-noise (S/N) in the shortest-wavelength
echelle orders, but at longer wavelengths the data give a good example
of some of the features present in massive stars in the CUBES range.
In practice, degrading the spectrum to the $R$\,$\sim$\,7,000 of the
low-resolution mode has little qualitative impact on the final
spectrum shown in Fig.~\ref{hd269896} given the (astronomical)
broadening of the lines.

Classified as ON9.7~Ia$+$ (and with log($L/L_\odot$)\,$\approx$\,6)
the UVES data of HDE\,269896 reveal the high Balmer series and a
plethora of He~\1 lines, together with weak emission lines from Si~\3
(\lam3487, 3590, 3807), Si~\4 (\lam3762 and 3773) and Al~\3 (\lam3602)
produced in its strong stellar wind (see \cite{ecfh04}). The presence
of lines from two ionisation stages of silicon offers a potential
temperature diagnostic in the CUBES region, albeit very dependent on
the adopted stellar parameters (including the wind) when the lines are
in emission.

To illustrate the ability of sophisticated model atmospheres and
spectral synthesis to reproduce this region in O-type spectra
(including the higher-order members of the Balmer series), in red in
Fig.~\ref{hd269896} we overplot the adopted {\sc cmfgen} model (see
\cite{cmfgen}) for this star from \cite{ecfh04}.  The adopted
parameters were an effective temperature, $T_{\rm
  eff}$\,$=$\,27,500~K, log$g$\,$=$\,2.7, and a rotational velocity of
$v$sin$i$\,$=$\,70\,\kms\ (see \cite{ecfh04} for further details of
the model parameters, which necessitated a detailed treatment of the
wind in this luminous supergiant star).

\section{Diagnostic lines in the near UV}

To investigate the sensitivity of different spectral lines in the near
UV to physical parameters such as temperature and gravity, we used
synthetic spectra from the extensive OSTAR2002
grid\footnote{http://tlusty.oca.eu/Tlusty2002/tlusty-frames-OS02.html
in which abundances in the OSTAR2002 grid are scaled relative to
solar values from \cite{gs98}.}  of line-blanketed, non-LTE,
plane-parallel, hydrostatic model atmospheres calculated with the {\sc
  tlusty} code \cite{tlusty}.

\subsection{LMC metallicity: Temperature trends}

The metallicity of the Large Magellanic Cloud (LMC) is approximately
half solar and we have a good understanding of stellar properties from
analysis of data at other wavelengths (e.g. \cite{vfts}). We therefore
first considered the 0.5\,$Z_{\odot}$ models from the OSTAR2002 grid
as example sub-solar models with which to investigate the CUBES domain.

We initially considered models with log$g$\,$=$\,4.0, that are
typical of the gravities estimated for dwarfs \cite{cssj17}. The {\sc tlusty}
grid has models from $T_{\rm eff}$\,$=$\,27,500 to 55,000\,K (in steps of 
2,500\,K). Using the effective temperature -- spectral type relations derived 
from analysis of O- and B-type dwarfs in the LMC \cite{cssj17,trundle07}, 
we can map the {\sc tlusty} models to approximate spectral types. 

\begin{figure}
  \includegraphics[width=12cm]{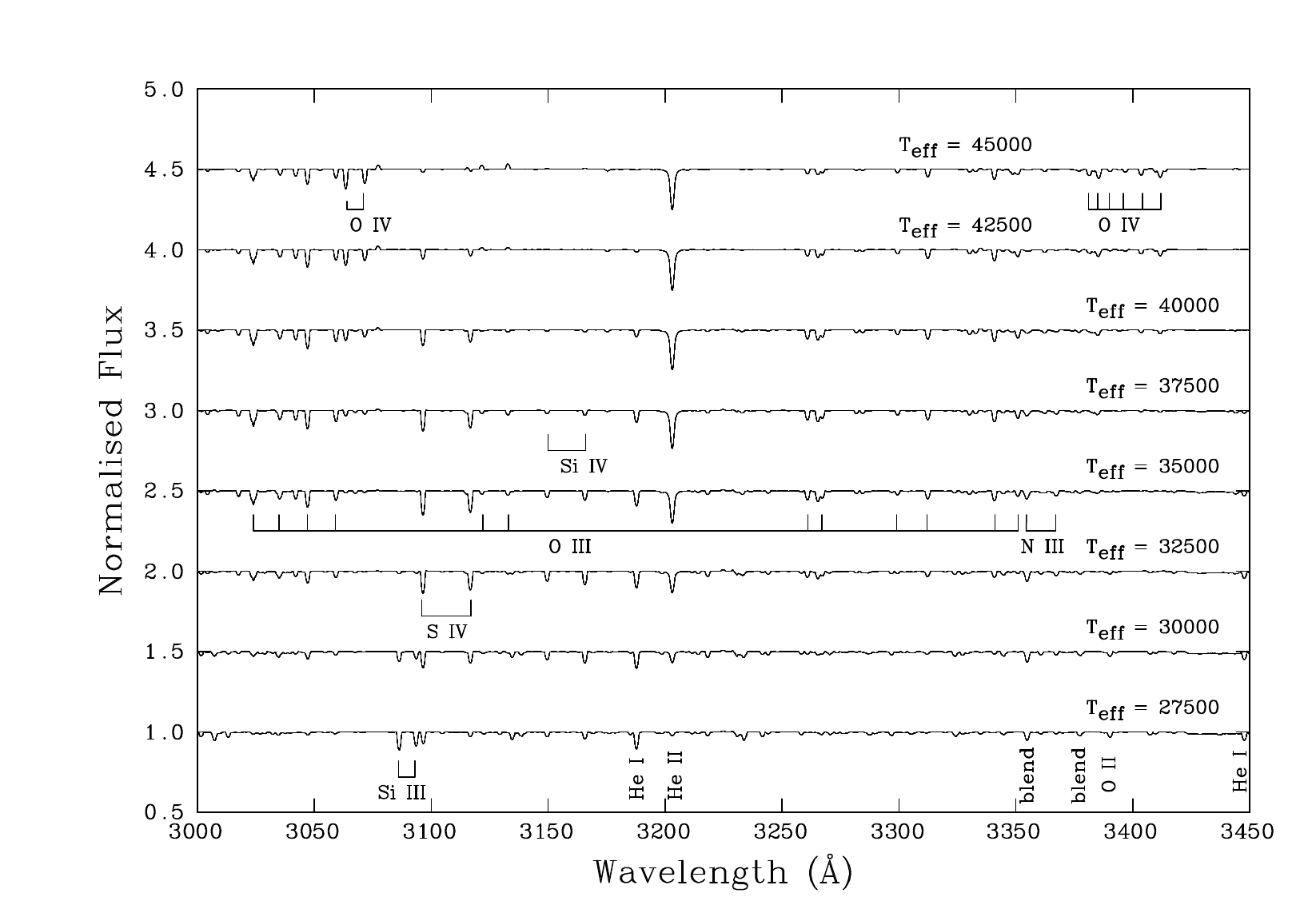}
  \caption{Synthetic near-UV spectra from the {\sc tlusty} $Z$\,$=$\,0.5$Z_{\odot}$ OSTAR2002 grid, which have been convolved with a rotational-broadening
    function of $v$sin$i$\,$=$\,100\,\kms. Identified lines are:
    He~{\scriptsize I} \lam\lam3188, 3448; He~{\scriptsize II} \lam3203; N~{\scriptsize III} \lam\lam3354, 3367 (where the former is blended with Ne~{\scriptsize II} 3355);
    O~{\scriptsize II} \lam\lam3390; O~{\scriptsize III} \lam\lam3024, 3035, 3047, 3059, 3122, 3133, 3261, 3267, 3299, 3312, 3341, 3351; O~{\scriptsize IV} \lam\lam
    3063, 3072; 3381, 3385, 3390, 3397, 3404, 3410-12-14; 
    Si~{\scriptsize III} \lam\lam3086, 3093; Si~{\scriptsize IV}
    \lam\lam3150, 3166; S~{\scriptsize IV} \lam\lam3097, 3118. Two blends are also identified at \lam3355 (He~{\scriptsize I} and Ne~{\scriptsize II}) and \lam3377 (O~{\scriptsize II} and Ne~{\scriptsize II}).}
\label{LMC_nuv}

  \includegraphics[width=12cm]{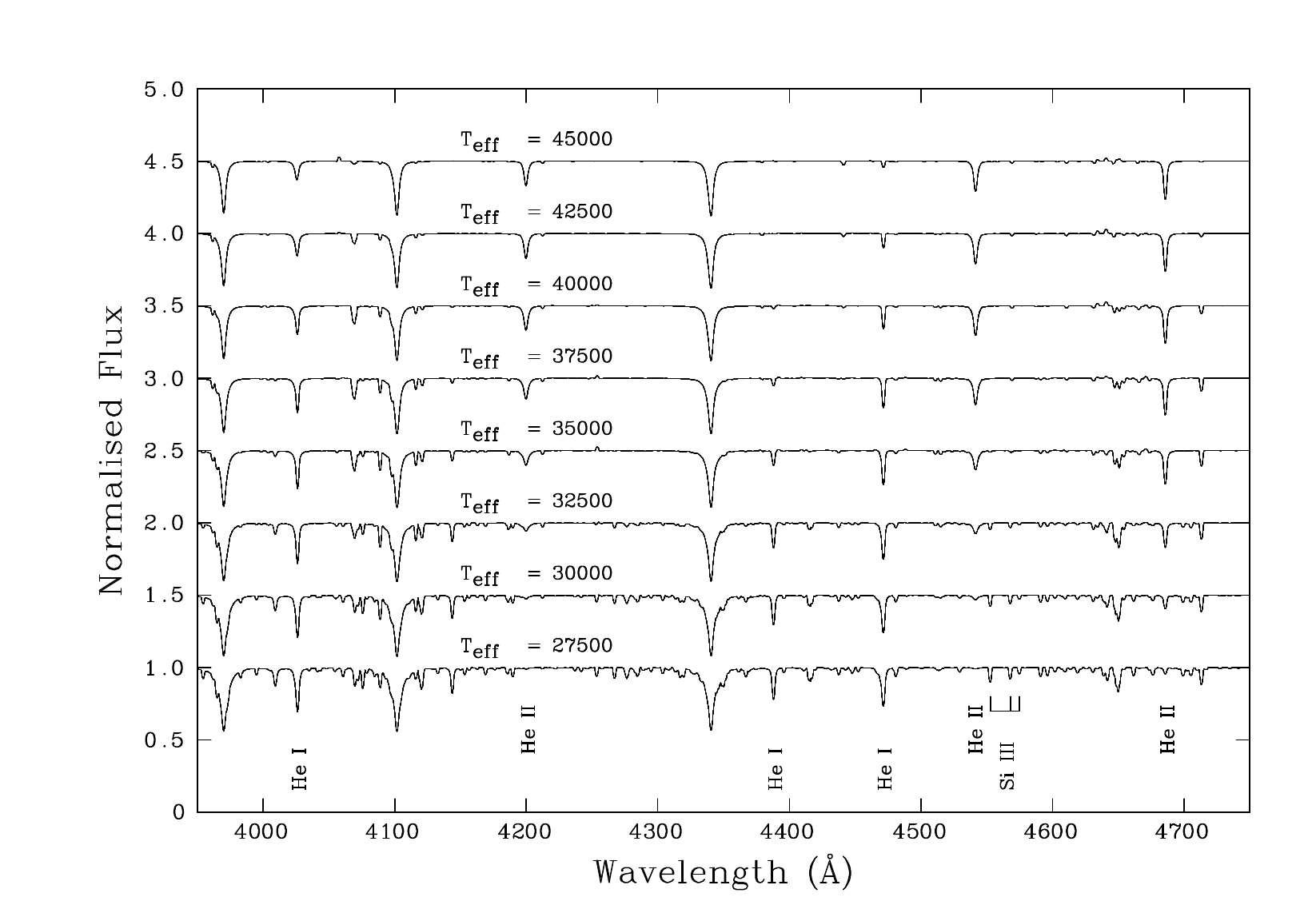}
  \caption{Synthetic spectra from the {\sc tlusty} $Z$\,$=$\,0.5$Z_{\odot}$ OSTAR2002 grid
    for the traditional classification domain. The
    spectra have been convolved with a rotational-broadening function
    of $v$sin$i$\,$=$\,100\,\kms. The identified lines are some of
    those used for classification of O-type spectra, namely: He~{\scriptsize I}
    \lam\lam4026 (blended with He~{\scriptsize II} at the hottest temperatures), 4388,
    4471; He~{\scriptsize II} \lam\lam4200, 4542, 4686; Si~{\scriptsize III} \lam\lam4552-68-75.}
\label{LMC_vis}

\end{figure}

The near-UV and visible ranges are
shown for eight model spectra in Figs.~\ref{LMC_nuv} and \ref{LMC_vis}, respectively. 
Each model spectrum in the figures has been convolved with a
rotational-broadening function of $v$sin$i$\,$=$\,100\,\kms\ to mimic
real targets compared to the unbroadened models. Note that
$v$sin$i$ of order 100\,\kms\ then dominates the broadening of the
line profiles, such that convolving the models to either the high- or
low-resolution modes of CUBES has a limited impact compared to the
spectra shown in the figure.

Our motivations to show the visible region of the models is twofold.
Firstly, to investigate if there are near-UV lines in the same models
that can then be used to delineate similar trends as in the temperature
(spectral type) sequence in Fig.~\ref{LMC_vis}.  Secondly, the models
shown illustrate some of the key classification criteria in O- and
early B-type stars (see \cite{wf00,sota11}). For instance, the
weakening of the He~\1 lines when moving to hotter temperatures
(earlier spectral types), such that the He~\1 \lam4471 line is very
weak by O3 ($T_{\rm eff}$\,$\approx$\,45,000\,K), and the weakening of
the He~\2 lines as we move to cooler temperatures (later spectral
types), such that the Si~\3 triplet is stronger than He~\2 \lam4542 by
B0. The $T_{\rm eff}$\,$=$\,40,000\,K model is also a useful
classification anchor point in terms of the He~\1/He~\2 ratios for the
\lam4026/\lam4200 and \lam4388/\lam4542 pairs, corresponding to a
classification of $\sim$O6.

To complement the UVES data of HDE\,269896, in Fig.~\ref{LMC_nuv} we
show the near-UV (\lam\,$<$\,3450\,\AA) spectra for the same models as
shown in Fig.~\ref{LMC_vis}.  The primary features of interest, in
terms of their potential usefulness for spectral classification and
determination of physical parameters are the He~\1 \lam3188 and He~\2
\lam3203 lines (previously studied by \cite{m75}).  The He~\1 line is
absent for the hottest spectra and increases in strength down the
temperature sequence, while the He~\2 changes in the opposite
sense. The lines are closest to being equivalent in the $T_{\rm
  eff}$\,$=$\,32,500 model, which would be classified as approximately
O9.7~V from comparison of its blue-visible spectrum
(Fig.~\ref{LMC_vis}) to published spectral standards \cite{sota11},
although the point at which the line intensities are equal is slightly
cooler.

This region is also rich with O~\3 absorption lines, as well as pairs
of lines from Si~\3 and Si~\4 and several O~\4 lines in the hottest
spectra. As at the longer wavelengths shown in Fig.~\ref{hd269896},
the presence of two ionisation stages in this region (from multiple
species) offers alternative potential diagnostics of temperature.

In this first exploration, with relatively modest rotational
broadening ($v$sin$i$ $=$\,100\,\kms), we have followed the general
classification approach for digital spectral of considering the central line
depths rather than the line intensities (equivalent widths).  Once
empirical data is available for a broad range of massive stars in this
spectral range, a more robust morphological treatment will be required in the
future, which e.g. takes into account effects such as rotational
broadening (e.g. \cite{markova11}, see also the discussion by \cite{arias16}).

\subsection{LMC metallicity: Stellar gravities}

The \lam3188/\lam3203 line ratio looks potentially interesting in the
context of (approximate) spectral classification and estimates of
$T_{\rm eff}$, but if we were limited to CUBES observations alone of a
given target we must also consider how to constrain its gravity (as
well as the possible impact that it has on the \lam3188/\lam3203
ratio). As demonstrated in Fig.~\ref{hd269896}, the CUBES range
contains the high Balmer series lines (plus the H8 and H$\epsilon$
lines not shown in the figure). The profile wings of the Balmer series
are generally excellent diagnostics of stellar gravity, and the high
Balmer lines could be used to constrain the gravities of CUBES
targets. With crossed-dispersed instruments such as UVES and
X-Shooter, correction of the echelle blaze function and stitching
together the different echelle orders can be challenging in the
3650-3900\,\AA\ region, where the wings of the Balmer lines can
overlap and it is difficult to accurately define a continuum. An
advantage of the CUBES design is the continuous spectrum
from each of its two arms, provided by having only one dispersing
element (in each arm) operating in first order \cite{zanutta22}, i.e. no echelle
orders to combine in the data reduction.

\begin{figure}
  \includegraphics[width=12cm]{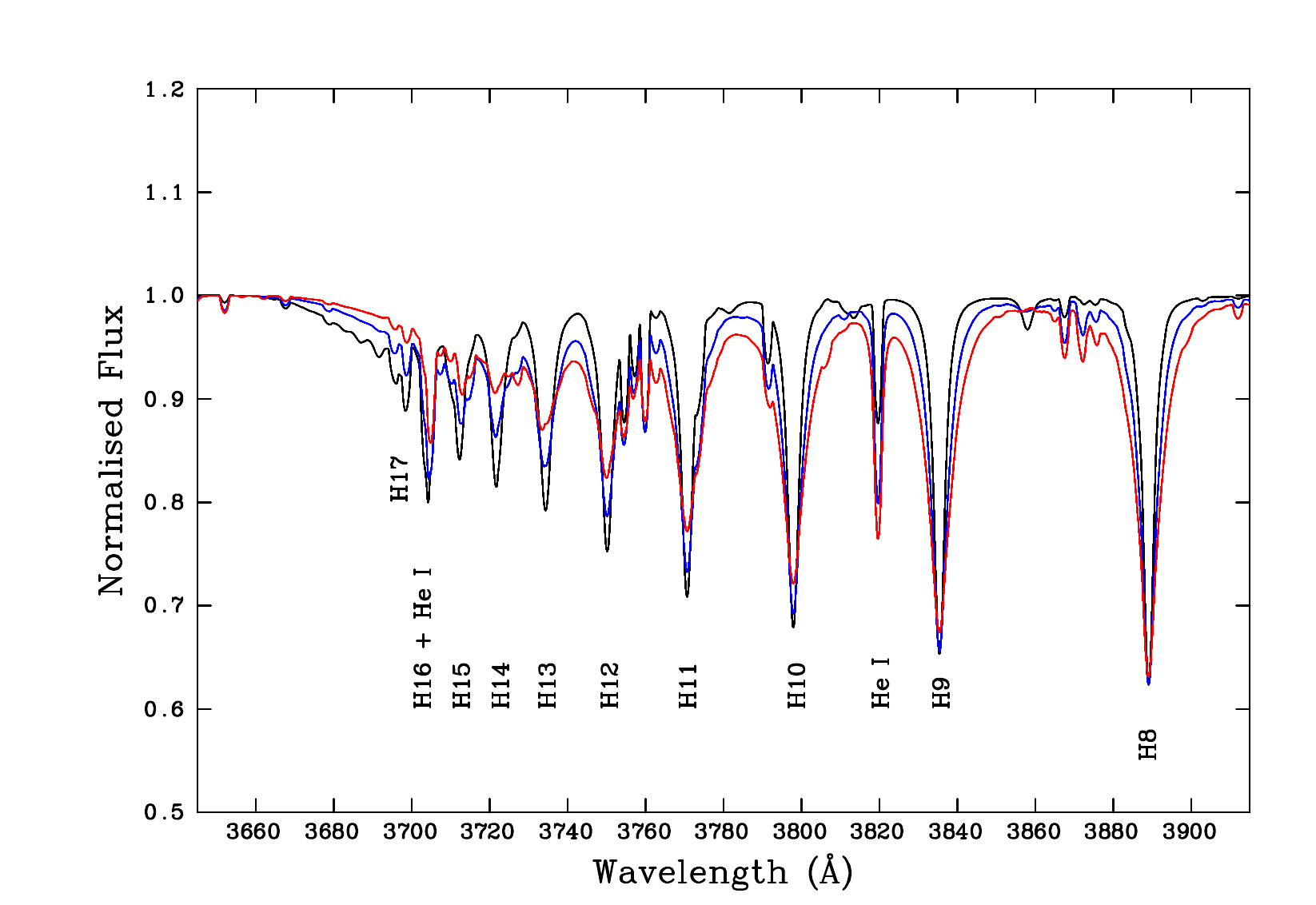}
  \caption{Synthetic spectra from the {\sc tlusty} $Z$\,$=$\,0.5$Z_{\odot}$ OSTAR2002 grid
    for the high Balmer lines for $T_{\rm
      eff}$\,$=$\,35,000 and three gravities: log$g$\,$=$\,3.5 (black
    spectrum), 4.0 (blue) and 4.5 (red). Identified lines are the
    same as those in Fig.~\ref{hd269896}, plus the H8 line
    (\lam3889).}
\label{LMC_logg}
\end{figure}

To illustrate the diagnostic potential of the Balmer lines in this
regard, in Fig.~\ref{LMC_logg} we show models for $T_{\rm eff}$\,$=$\,35,000\,K (typical of a late O-type star) for three
gravities: log$g$\,$=$\,3.5 (a typical value for an O-type giant in the LMC, e.g. \cite{ora17}), 4.0 and 4.5, spanning from shortwards of
the Balmer limit up to the H8 line. We recognise the challenges of the
blending of the higher-series lines (particularly for noisy data), but
note that the H8 line is relatively isolated, and the CUBES range
(which extends further redwards to 4050\,\AA) also includes the
H$\epsilon$ line. While the latter is blended with the interstellar
Ca~\2 $K$ line at \lam3968, its redward wing could be used to
provide further constraints.

\begin{figure}
\begin{center}
  \includegraphics[width=10cm]{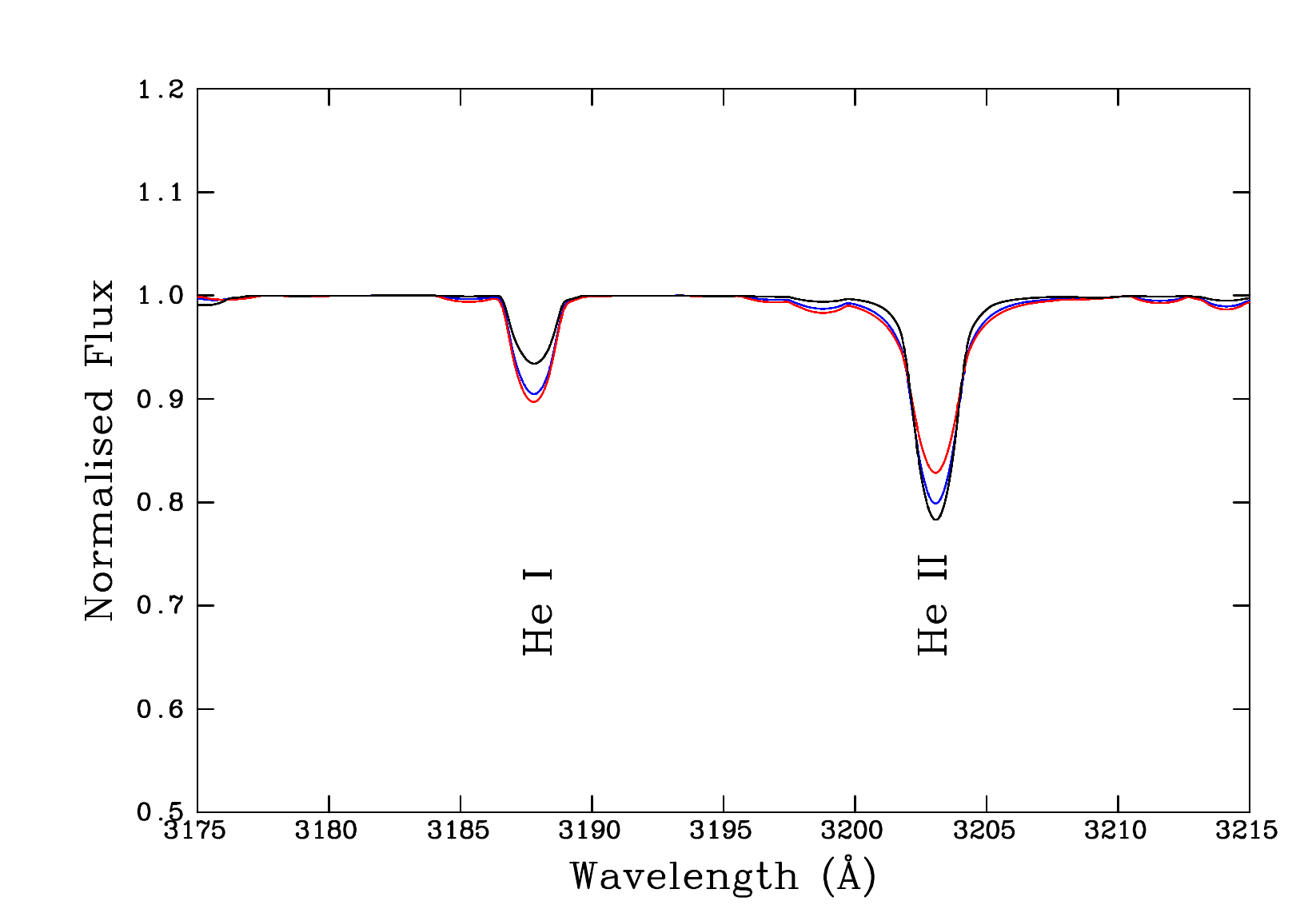}
  \caption{Synthetic {\sc tlusty} spectra ($Z$\,$=$\,0.5$Z_{\odot}$)
    for the He~{\scriptsize I} \lam3188 and He~{\scriptsize II}
    \lam3203 lines for $T_{\rm eff}$\,$=$\,35,000 and three gravities:
    log$g$\,$=$\,3.5 (black spectrum), 4.0 (blue) and 4.5 (red).}
\label{LMC_nuv_logg}
\end{center}
\end{figure}

The adopted gravity does have an impact on the strengths of the He~\1
\lam3188 and He~\2 \lam3203 lines (as shown in Fig.~\ref{LMC_nuv_logg}), with the
former being stronger and the latter being weaker at higher gravities.
As such, good S/N across the full CUBES range will be critical to
ensure the maximum information is available, but the key point is that
there are potential diagnostics available of both temperature and
gravity.

A caveat of our approach in using the {\sc tlusty} models is that the
effects of stellar winds are not included, which would be expected to
modify the appearance of the emergent spectra. However, we note that
the winds of O-type giants and dwarfs at sub-solar metallicity are generally weak,
and the {\sc tlusty} models are sufficient for our qualitative
objectives here.

\subsection{Low-metallicity models ($Z$\,$=$\,1/30\,$Z_\odot$)}
\label{sec:2}

The LMC-like models described above will be relevant for future CUBES targets in external galaxies, but the real push is to extend studies to lower metallicities than currently possible (see \cite{voyage2050}). A significant effort over the past 20 years has gone into quantifying the impact that metallicity has on the evolution of massive stars, so that we can improve stellar-evolution and population-synthesis models to more accurately reproduce the massive-star populations seen in both the local Universe and high-redshift, star-forming galaxies. Necessarily, most efforts have focused on massive stars in the Milky Way, LMC and the Small Magellanic Cloud (SMC), spanning a range of metallicities from that of the solar neighbourhood down to approximately one-fifth solar in the SMC. This puts limits on our ability to test model predictions at lower metallicities, and requires (uncertain) extrapolations.

High-efficiency, multi-object spectrographs, such as the FOcal Reducer/low dispersion Spectrograph~2 (FORS2) on the VLT, have provided first insights (at $R$\,$\sim$\,1,000) into limited numbers of massive stars in more distant systems, with nebular oxygen abundances of $\sim$0.15$Z_{\odot}$, namely: IC\,1613 at 0.7\,Mpc, WLM at 0.9\,Mpc, and NGC\,3109 at 1.2\,Mpc. However, a key motivation to improve our understanding of the physical properties of massive stars, and their contribution to galaxies in the early Universe, is to directly observe stars at even lower metallicites. Example targets in this context include: Sextans~A (0.1$Z_{\odot}$ at 1.3\,Mpc), SagDIG (0.05$Z_{\odot}$ at 1.1\,Mpc), Leo~P (0.03$Z_{\odot}$ at 1.6\,Mpc) and, ultimately, I\,Zw18 (0.02$Z_{\odot}$ at 18.9\,Mpc).

\begin{figure}
  \includegraphics[width=12cm]{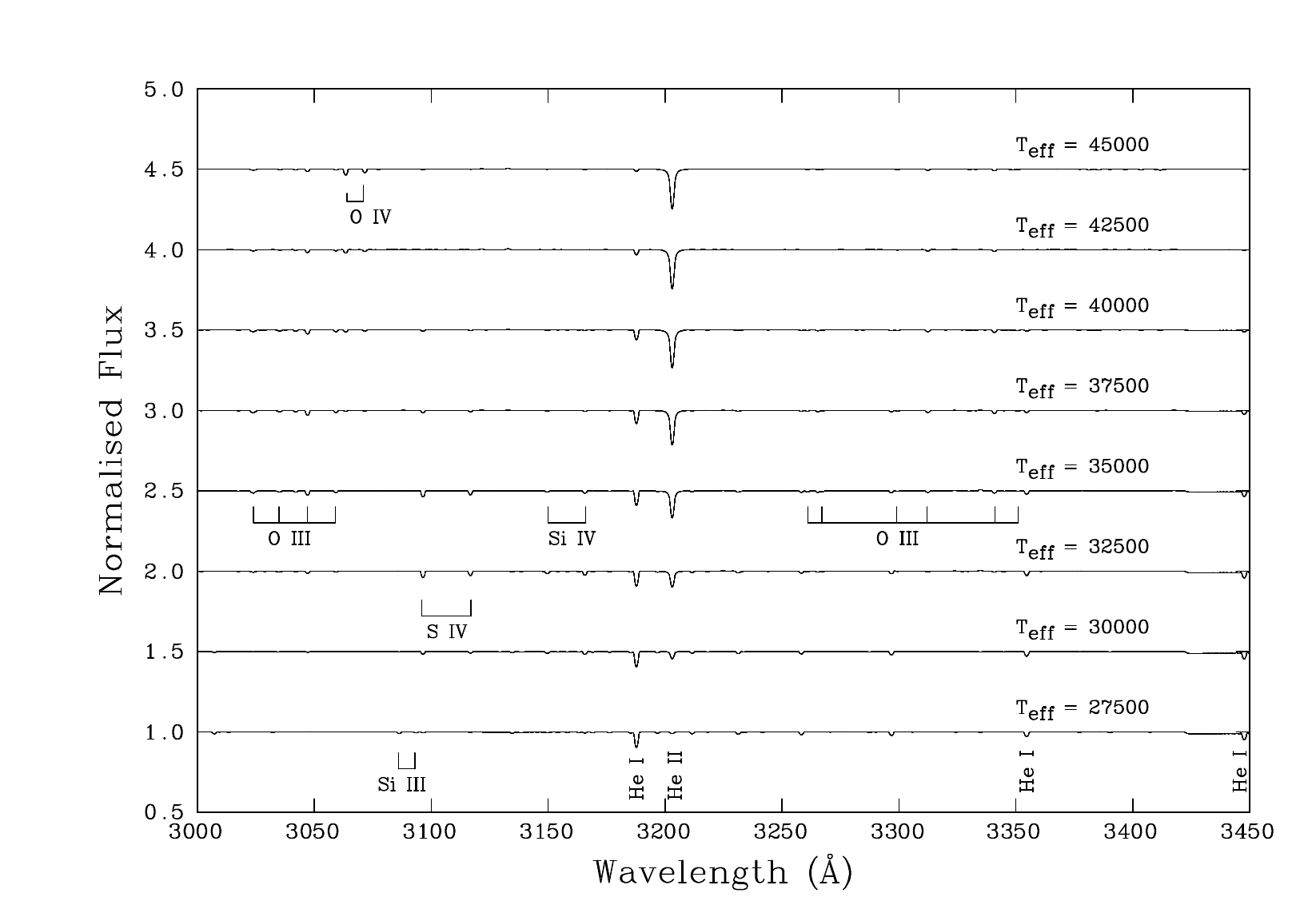}
  \caption{Synthetic near-UV spectra from the {\sc tlusty} $Z$\,$=$\,0.03$Z_{\odot}$ OSTAR2002 grid, which have been convolved with a rotational-broadening
    function of $v$sin$i$\,$=$\,100\,\kms. Identified lines are:
    He~{\scriptsize I} \lam\lam3188, 3355, 3448; He~{\scriptsize II} \lam3203;
    O~{\scriptsize III} \lam\lam3024, 3035, 3047, 3059, 3261, 3267, 3299, 3312, 3341, 3351; O~{\scriptsize IV} \lam\lam3063, 3072; Si~{\scriptsize III} \lam\lam3086, 3093; Si~{\scriptsize IV} \lam\lam3150, 3166; S~{\scriptsize IV} \lam\lam3097, 3118.}
\label{LeoP_nuv}

  \includegraphics[width=12cm]{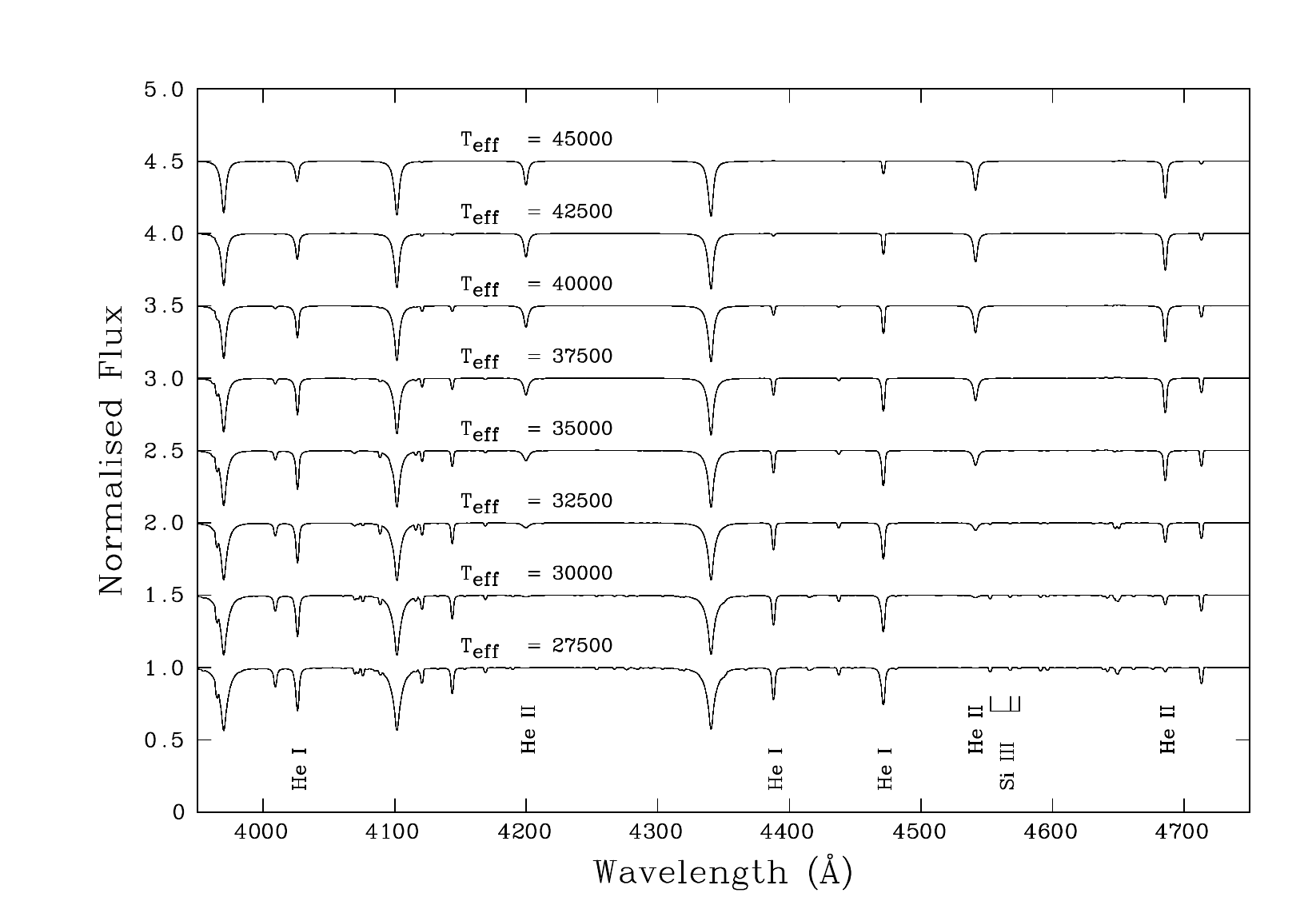}
  \caption{Synthetic spectra from the {\sc tlusty} $Z$\,$=$\,0.03$Z_{\odot}$ OSTAR2002 grid
    for the traditional classification domain. The
    spectra have been convolved with a rotational-broadening function
    of $v$sin$i$\,$=$\,100\,\kms. Identified lines are the same as in Fig.~\ref{LMC_vis}.}
\label{LeoP_vis}
\end{figure}

To investigate the spectral lines in such extremely metal-poor stars,
we have used the 0.03$Z_{\odot}$ models from the {\sc tlusty}
OSTAR2002 grid. As a first comparison with the LMC-like models, the
near-UV and visible regions for the 0.03$Z_{\odot}$ models are shown
in Figs.~\ref{LeoP_nuv} and \ref{LeoP_vis}, respectively. Echoing the
shift to higher temperatures for a given zero-age main-sequence mass
at lower metallicity (e.g.  \cite{IZw18}, and references therein),
note the He~\1/He~\2 line ratios in the $T_{\rm eff}$\,$=$\,40,000\,K
model compared to that in Fig.~\ref{LMC_vis}. The $T_{\rm
  eff}$\,$=$\,40,000\,K model for 0.03$Z_{\odot}$ would be classified
as a slightly later type than that for the 0.5$Z_{\odot}$ model.
Similarly, the He~\1 \lam4471 line is stronger in the 0.03$Z_{\odot}$
model.  In short, a spectrum classified as, e.g. O3 or O6, at very low
metallicity (e.g. in Leo~P or I\,Zw18) would have a hotter temperature
than its morphological counterpart in the LMC.

The 3000-3450\,\AA\ near-UV region for the 0.03$Z_{\odot}$ models is shown in Fig.~\ref{LeoP_nuv}. As expected, the metallic lines are now significantly weaker, but there are still several weak O~\3 lines present in the models corresponding to later O-types ($T_{\rm eff}$\,$=$\,32,500 \& 35,000). As with the He~\1/He~\2 line ratios in the visible 
(Fig.~\ref{LeoP_vis}), there is a shift in temperature of when the He~\1 \lam3188 and He~\2 \lam3203 lines have equivalent intensities at lower metallicity.  They are roughly equivalent at $T_{\rm eff}$\,$=$\,32,500 in the 0.03$Z_{\odot}$ models (Fig.~\ref{LeoP_nuv}), compared to somewhere between $T_{\rm eff}$\,$=$\,32,500 and 30,000 in the 0.5$Z_{\odot}$ models (Fig.~\ref{LMC_nuv}). Again this reflects the hotter models needed to reproduce a given line ratio at lower metallicity (i.e. the temperature for a given spectral type, on the basis of using Galactic morphological criteria, would be higher).

\section{Stellar properties in Local Group galaxies}

\subsection{Oxygen abundances}
The key properties of a (single) massive star are defined by its initial mass and
metallicity, but its path in the Hertzsprung--Russell diagram and ultimate fate depend
critically on both mass loss and rotation \cite{puls08,langer12}. Studies have shown that the effects of rotation are enhanced at low metallicity (e.g. \cite{mm01,brott11}), and rotationally-enhanced mixing is predicted to bring CNO-processed material to the stellar surface. 
To test the various implementations and prediction of rotation and mixing in stellar evolution models we require good empirical constraints of CNO abundances as a function of temperature, luminosity, metallicity and rotation rates.  

As illustrated in Fig.~\ref{LMC_nuv}, the near UV spectra of O-type
stars contain a plethora of oxygen lines, with several O~\4 lines in
the hottest spectra, a large number of O~\3 lines in mid-late types,
and weak O~\2 lines in the coolest spectra (e.g. \lam3390).  These
lines are the best available diagnostics of O abundances in massive
stars, because those at other wavelengths are generally saturated and
strongly affected by stellar winds, making abundance determination
both complicated and uncertain. Moreover, the use of multiple lines is
also critical to arrive at accurate abundances with reliable error
bars (see Fig.~1 from \cite{martins15}).  We add that accurate oxygen
abundances are required to use the O~\5 \lam1371 line from far-UV
observations to estimate stellar wind densities around the sonic point
and to explore the effect of clumping in the wind to arrive at
reliable mass-loss rates.  There are also useful lines from C and N in
the near-UV range (e.g. C~\3 \lam3609 and N III \lam\lam3354-67-74)
that appear to free of wind effects in relevant {\sc cmfgen} models
(e.g. \cite{mb22}), but these warrant further investigation.

\begin{figure}
  \includegraphics[width=12cm]{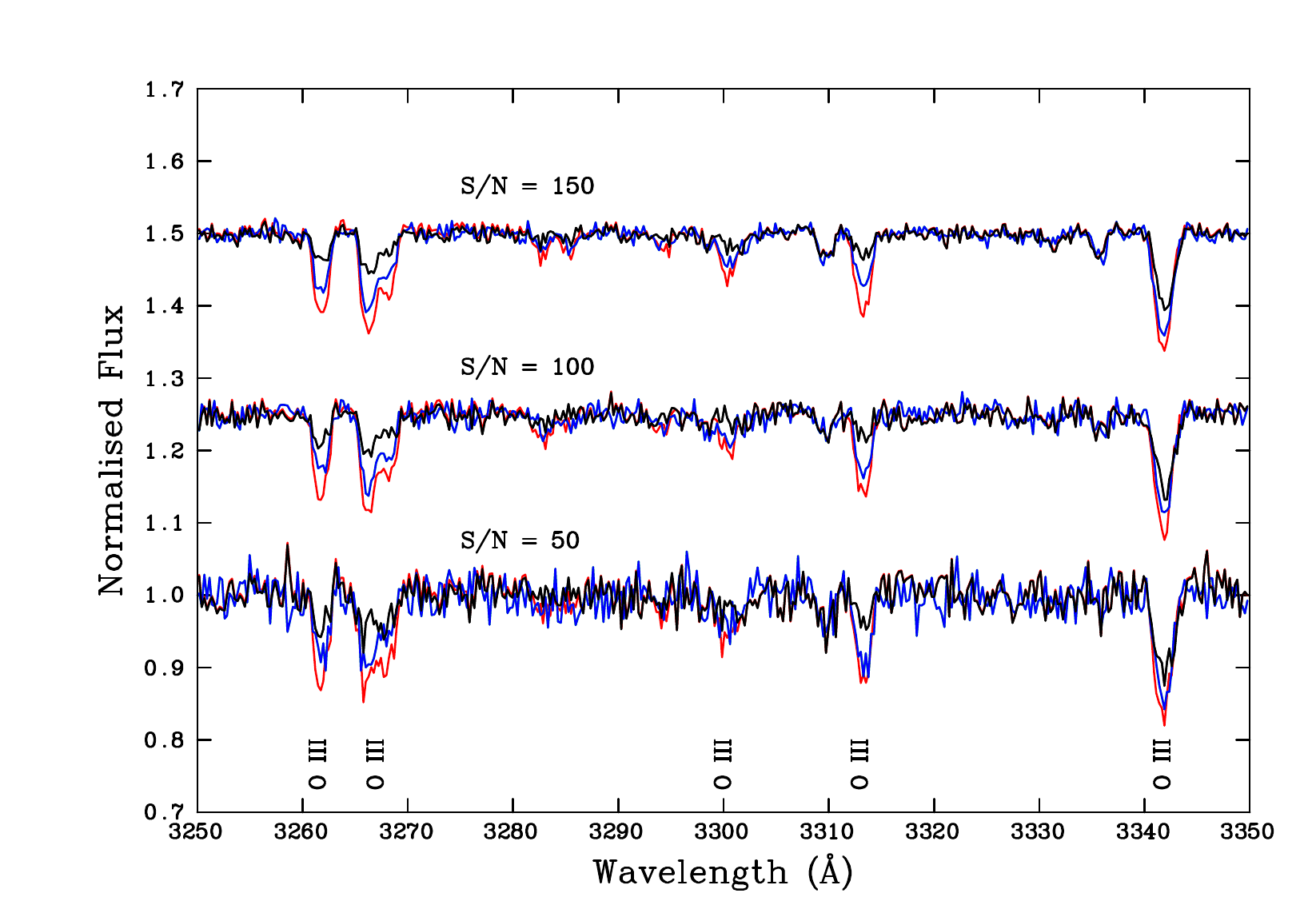}
  \caption{The impact of noise on the O~{\scriptsize III} lines in {\sc cmfgen} model spectra of an O8-type star in the LMC. Three oxygen abundances are shown, a LMC model with 12\,$+$\,log(O/H)\,$=$\,8.39 (blue), and models with 12\,$+$\,log(O/H)\,$=$\,8.74 (red) and 7.78 (black, comparable to results in WLM and NGC\,3109) to illustrate the sensitivity of the lines. Identified lines are O~{\scriptsize III} \lam\lam3261, 3267 (blend), 3299, 3312, 3341. Noise has been added to the models to simulate different signal-to-noise (S/N) ratios in the continuum as indicated.}
\label{O_models}
\end{figure}

To illustrate the sensitivity of the O~\3 lines to abundance changes and the S/N of the observations, in Fig.~\ref{O_models} we show a {\sc cmfgen} model \cite{cmfgen} with T$_{\rm eff}$\,$=$\,31,000\,K and log($g$)\,$=$\,3.1 for HD\,269702 (classified as O8 I(f)p, see \cite{walborn2010}) in the LMC. The baseline oxygen abundance is 12\,$+$\,log(O/H)\,$=$\,8.39 (shown in green), compared with models for 12\,$+$\,log(O/H)\,$=$\,8.74 (in red) and 7.78 (comparable to results for metal-poor irregulars, in black).
We then introduced model noise to each spectrum to reproduce continuum S/N levels of 50, 100 and 150 as shown in the figure.

As expected given past observational studies (e.g. \cite{martins15}), the spectra in Fig~\ref{O_models} suggest that S/N\,$\gtrsim$\,100 is required to estimate O abundances in the LMC and more metal rich targets as the lines becomes saturated. For instance, there is little sensitivity to changes in abundance over this abundance range in the O~\3 \lam3312 line. However, the situation is less challenging at lower abundances -- although S/N remains critical, the lines are more responsive to abundance changes. As noted above, the precision on the estimated O abundance can also be improved by analysing multiple lines together (recalling the large number of O~\3 lines in the mid-late O-type spectra in Fig.~\ref{LMC_nuv}, and the O~\4 lines in the hottest spectra).

To date, estimates of O abundances in B-type supergiants have been possible in metal-poor galaxies at the fringes of the Local Group, e.g. 12\,$+$\,log(O/H) $=$\,7.83\,$\pm$\,0.12 in WLM \cite{bresolin06} and 7.76\,$\pm$\,0.07 in NGC\,3109 \cite{evans07}. Similar techniques were also used to estimate light-element abundances (C, N, O, Si, Mg) of B-type supergiants at $\sim$2\,Mpc in NGC\,55 \cite{castro55}. However, while the supergiants are a useful reference point for the evolutionary models (and for comparisons with nebular abundances), for true insights into the physical processes on the main sequence, we need similar studies of O-type stars. Even though we have confirmed O stars in WLM and NGC\,3109 \cite{bresolin06,evans07}, as well as in IC\,1613 \cite{garcia13} and Sextans~A \cite{Cal16}, abundance estimates are out of reach of present observations (e.g. \cite{tramper11,tramper14}).

\subsection{Stellar rotational velocities}

Another important observational property of massive stars is their projected rotational velocity ($v$sin$i$). As noted above, stellar rotation is important in the context of chemical enrichment (e.g. CNO abundances), and observational estimates of $v$sin$i$ are also powerful diagnostics of rapidly-rotating stars that have experienced chemically-homogeneous evolution or that have previously undergone binary interaction.

The helium lines in visible spectra are often used to estimate rotational velocities of O-type stars (e.g. \cite{rss13}). Although they are affected by Stark broadening (typically with an equivalent FWHM of 50-100\,\kms), the helium lines can be used to study the overall distribution of projected rotational velocities for samples of stars, while also allowing identification of rapid rotators. However, working with low-resolution ($R$\,$\sim$\,1000) spectroscopy from, e.g. FORS2 or OSIRIS (on the Gran Telescopio Canarias), at S/N\,$\sim$\,50 we are unable to put even modest constraints on $v$sin$i$; the minimum combination required is $R$\,$>$\,2000 and S/N\,$>$\,100.

A more robust probe of rotational velocities is provided by metallic lines in the spectra of massive stars. The metallic lines do not suffer from Stark broadening, nor do they suffer nebular contamination which can also hamper the use of helium lines in extragalactic targets. The visible spectra of B-type stars are replete with isolated metal lines that can be used to investigate $v$sin$i$ (e.g. Si~\3, Si~\4, Mg~\2 etc) but the only comparably useful probe for O-type stars is the (less commonly observed) O~\3 \lam5591 line. 

The capabilities of CUBES are particularly compelling in this context. The combination of its sensitivity, improved spectral resolution (compared to e.g. FORS2) and access to the broad range of metallic lines in the near UV (see Fig.~3) will enable robust estimates of $v$sin$i$ for metal-poor stars at the edge of the Local Group. High S/N observations would also enable investigation of the contribution of macroturbulent broadening in (sub-SMC) metal-poor O-type stars for the first time (e.g. \cite{a09,ssd10}). For instance, for stars with relative narrow lines, S/N\,$\gtrsim$\,100 is sufficient to use Fourier transform analysis to estimate the contribution of macroturbulence \cite{ssd14}.

\subsection{Alternative temperature diagnostics}

Beyond the O-type stars discussed here, the coverage of the Balmer jump provided by CUBES could also provide valuable constraints on effective temperature for other extragalactic targets, as used for e.g. A-type supergiants \cite{kud08}, B-type supergiants \cite{z09}, and Be-type stars \cite{s18}. This technique typically requires accurate determination of the flux levels at either side of the Balmer limit, rather than precise absolute flux calibration. We do not explore this application further here, but note it for completeness and as an example of where flux calibration of the spectra will be important to have a good understanding of the wavelength-dependent properties of the spectra (response function, slit-losses etc).

\section{CUBES Performances}
In contrast to lower-mass stars, the spectral energy distributions of
massive stars peak in the far-UV. This means that observations
with CUBES will probe the rising part of the flux distribution,
potentially opening-up observations of targets that are otherwise too
faint to observe with other facilities.  For instance, the intrinsic
$U-V$ colour for a mid O-type star is $(U-V)_0$\,$\sim$\,$-$1.5\,mag
(e.g. \cite{johnson}), representing a potentially significant gain
compared to observations at visible wavelengths.

One caveat to this potential gain is the challenge of
line-of-sight extinction towards potential targets, as its effects
become more significant at shorter wavelengths (with
$A(U)$\,$\approx$\,1.5\,$A(V)$ \cite{cardelli}).  Nonetheless, most
potential extragalactic targets are sufficiently far from the Galactic plane,
such that foreground extinction will not be too much of a limiting
factor, although targets in external galaxies might have to be
carefully selected to avoid those with a significant local
contribution to the line-of-sight extinction.

To estimate the potential performance of CUBES in the studies of
O-type stars in galaxies such as IC\,1613, WLM and NGC\,3109 we used some of the model spectra
discussed above as inputs to the Exposure Time Calculator (ETC)
developed during the conceptual design phase \cite{genoni22}.
The confirmed O-type stars in these irregular galaxies have $V$\,$=$\,19 to 20.5\,mag.
The S/N predictions from the ETC for 2\,hr exposures of three of the {\sc tlusty} 0.5$Z_{\odot}$ dwarf models are summarised in Table~\ref{sn_results}; adopted parameters for each calculation were: airmass\,$=$\,1.2 (a reasonable assumption for the example galaxies), seeing\,$=$\,0.8$''$, spatial binning\,$\times$2, spectral binning\,$\times$4. 

Given the similarity in the spectral energy distributions, the same ETC calculations using the 0.03$Z_{\odot}$ are nearly identical in the predicted S/N, so the use of the LMC-like models here does not unduly influence the results. The S/N values are quoted for two wavelengths: 3195\,\AA\ as representative of the continuum S/N near the He~\1 \lam3188 and He~\2 \lam3203 lines, and at 3640\,\AA\ to indicate the S/N shortwards of the Balmer limit.

The image slicer for the low-resolution mode generates a wider effective slit such that the resulting spectra are oversampled (with $>$9\,pixels per resolution element), hence the adopted $\times$4 spectral binning in the calculations to bolster the resulting S/N without loss of resolution. Equally, for the faintest massive stars, we could bin by a further factor of two to improve the S/N (i.e. $\sqrt{2}$\,$\times$\,S/N$_{\rm ETC}$), degrading the spectrum to an effective resolving power of $R$\,$\sim$\,3,500 (still more than three times that obtained for 3\,hr integrations of massive stars beyond 1\,Mpc with FORS2, e.g. \cite{evans07}). In the low-resolution mode the slicer has six 1$''$ slices, so in the ETC calculations we only extracted the central two slices (to optimise the S/N).

With additional binning spectrally, the results in Table~\ref{sn_results} demonstrate that it should be possible to obtain sufficient S/N ($\gtrsim$\,100) for studies of the physical parameters and oxygen abundances of O-type stars in galaxies at the edges of the Local Group (i.e. 1.2-1.3\,Mpc), with slightly longer ($\sim$2.5-3\,hr) exposures of the fainter known O stars in these systems.

\begin{table}
\begin{center}
\caption{Signal-to-noise (S/N) predictions for CUBES observations of massive stars with the low-resolution mode for illustrative LMC-like ($Z$\,$=$\,0.5$Z_{\odot}$) {\sc tlusty} models. Adopted parameters for the Exposure Time Calculator (ETC) are: exposure time\,$=$\,2\,hr, airmass\,$=$\,1.2, seeing\,$=$\,0.8$''$, spatial binning\,$\times$2, spectral binning\,$\times$4 (given that the low-resolution mode is sampled with $>$9 pixels per resolution element).}
\label{sn_results}
\begin{tabular}{ccccccc}
\hline\noalign{\smallskip}
 $V$ & \multicolumn{2}{c}{$T_{\rm eff}$\,$=$\,45.0\,kK} & \multicolumn{2}{c}{$T_{\rm eff}$\,$=$\,37.5\,kK} & \multicolumn{2}{c}{$T_{\rm eff}$\,$=$\,27.5\,kK} \\
\noalign{\smallskip}
(mag.)  & S/N$_{\lambda3195}$ & S/N$_{\lambda3640}$ & S/N$_{\lambda3195}$ & S/N$_{\lambda3640}$ & S/N$_{\lambda3195}$ & S/N$_{\lambda3640}$ \\ 
\noalign{\smallskip}\hline\noalign{\smallskip}
19.0 & \po95 &   113 & \po93 &   111 & \po86 &   104 \\
19.5 & \po74 & \po88 & \po72 & \po86 & \po66 & \po81 \\
20.0 & \po57 & \po67 & \po55 & \po66 & \po50 & \po62 \\
20.5 & \po43 & \po51 & \po42 & \po50 & \po38 & \po47 \\
\noalign{\smallskip}\hline
\end{tabular}
\end{center}
\end{table}

\subsection{`ELT science' with the VLT}

As an exciting example of where CUBES can bring observations into reach that are beyond our current capabilities, we highlight the case of star LP~26 in the dwarf Leo~P galaxy
\cite{leoP}. Deep (8\,hr) observations with the Multi-Unit Spectroscopic Explorer (MUSE) instrument revealed weak He~\2 \lam\lam4686, 5411 absorption lines, providing the first direct evidence of an O-type star in Leo~P. This is particularly interesting as the
significantly low oxygen abundance (3\% solar) of the H~\2 region in Leo~P suggests it has a near-primordial composition \cite{skillman13}. 

Although relatively nearby (1.6\,Mpc \cite{mcquinn}) for such a metal-poor system, Leo~P is sufficiently far away that further visible spectroscopy of its hot stars is beyond our current capability. For instance, the {\em HST} magnitudes for LP~26 are $F475W$\,$=$\,21.5 and $F814W$\,$=$\,21.8 (which, taking these as proxies for $V$ and $I$ filters compared to the anticipated intrinsic colours for O-type stars, suggest a low line-of-sight extinction). This puts it beyond the reach of (feasible) observing proposals with existing visible spectrographs, and further visible spectroscopy of the stellar population of Leo~P was thought to have to wait until the Extremely Large Telescope (ELT) is operational. 

First constraints on the properties of LP~26 have recently been provided by analysis of far-UV {\em HST} spectroscopy, reporting a fast rotational velocity ($v$sin$i$\,$=$\,370\,$\pm$\,90\,\kms) and, as expected given the lower metallicity, weaker wind lines than in comparison templates of SMC stars \cite{telford21}. Using the same OSTAR2002 {\sc tlusty} grid as in our calculations, the far-UV spectroscopy was combined with multiband photometry to estimate its temperature as $T_{\rm eff}$\,$=$\,37,500\,K, with uncertainties of order $\pm$6\,kK. 

The near-UV lines identified in the CUBES range offer the prospect of characterising the physical properties of LP~26. We used the 3\% solar $T_{\rm eff}$\,$=$\,37,500\,K {\sc tlusty} model to estimate the performance for $V$\,$=$\,21.5\,mag. with the ETC. The same parameters were used as in Table~\ref{sn_results}, including airmass\,$=$\,1.4, as this corresponds to the maximum altitude that Leo~P (with a declination of $+$18$^\circ$) reaches from Paranal. Extracting only the central two slices again, the ETC predicts S/N\,$=$\,34 at 3640\,\AA\ in a total integration of 4\,hrs; binning this (spectrally) by a further factor of two would provide S/N\,$\sim$\,50 in a relatively modest half night of observations; the same calculation for 3195\,\AA\ (central slice only, binning the ETC results by a further factor of two yields S/N\,$\sim$\,40. 

Such an observation would enable estimates of the physical parameters (temperature, gravity) of LP~26 in a relatively modest amount of observing time (half a night). This nicely illustrates the potential of CUBES in this scientific area -- we expect more candidate O-type stars to be discovered in the coming years at 1~Mpc and beyond, and CUBES could provide a powerful capability to constrain their physical parameters.

\begin{figure}
  \includegraphics[width=12cm]{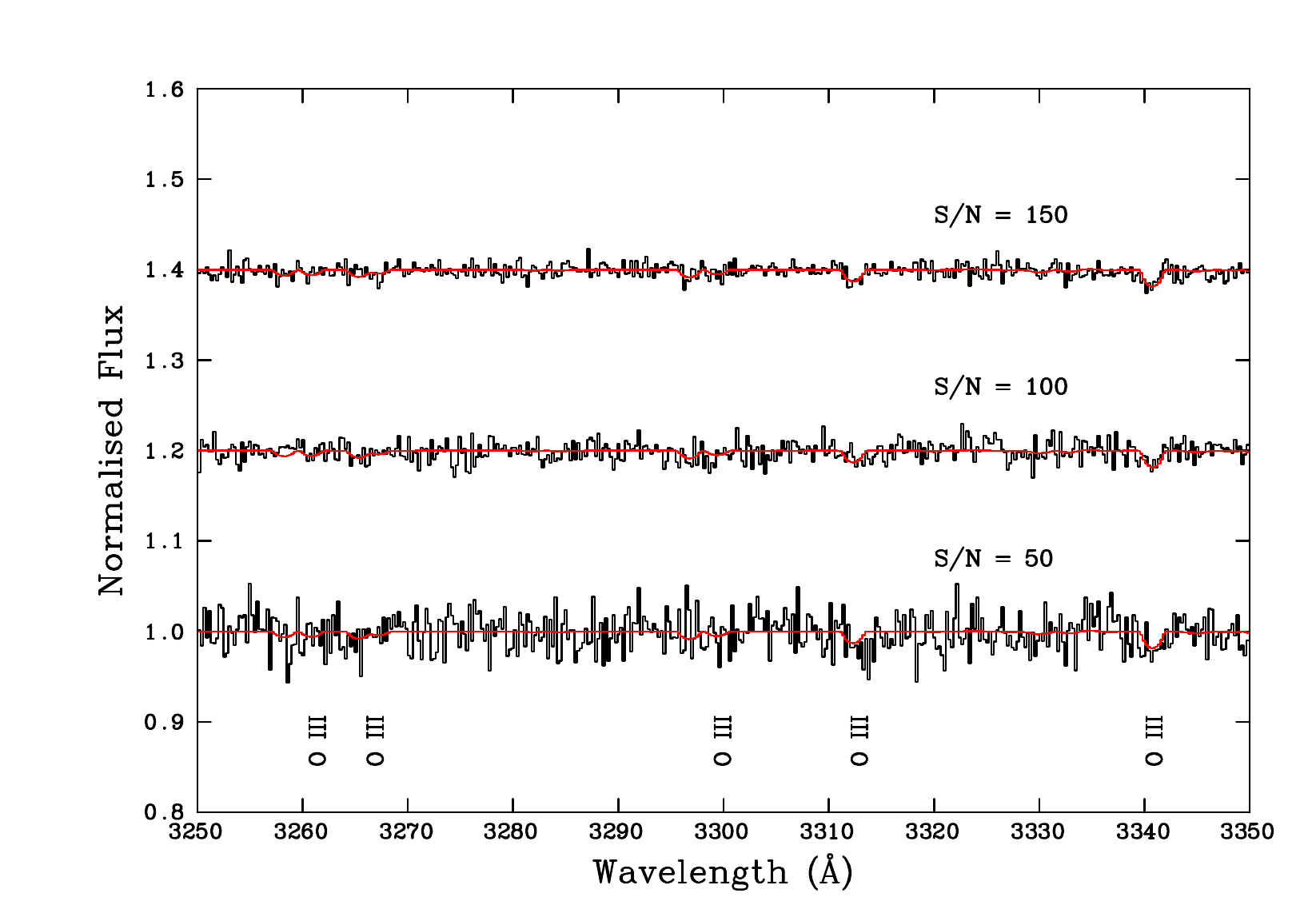}
  \caption{Synthetic {\sc tlusty} spectra ($Z$\,$=$\,0.03$Z_{\odot}$, $T_{\rm eff}$\,$=$\,37,500, log$g$\,$=$\,4.0) to investigate the potential of recovering oxygen abundances for different levels of simulated signal-to-noise (S/N) in the test case for Leo~P (see Section~5). The spectra with added noise are shown in black, with the original model spectrum overplotted on each in red. The identified O~{\scriptsize III} lines are the same as those in Fig.~\ref{O_models}.}
\label{O_models_lowZ}
\end{figure}

Direct determination of the oxygen abundance at the low metallicity of Leo~P will, however, remain challenging. The metal lines are sufficiently weak that greater S/N ($>$\,100) is required for secure detections. To explore this further, in Fig.~\ref{O_models_lowZ} we show the same {\sc tlusty} model spectrum for the O~\3 lines in the 3250-3350\,\AA\ region. A continuum S/N in excess of 100 is required to tentatively detect the O~\3 \lam3341 line, with S/N\,$=$\,150 being a more realistic goal if equivalent widths (or even firm upper limits) were to be measured. From the ETC, recovering a S/N of order 150 at 3350\,\AA\ (assuming the same binning as above in the ETC, and a further factor of $\sqrt{2}$) would require a total integration of $\sim$45\,hr. Such an ambitious observation is unlikely to be feasible, particularly given e.g. systematics that might limit performance from combining a large number of exposures together, although we will re-assess this case later in the construction phase of CUBES. 

\subsection{Impact of O$_3$ absorption}

Shortwards of $\sim$3400\,\AA\ the extinction due to atmospheric ozone
(O$_3$) becomes a significant factor, with a steep dependence of the
extinction as a function of airmass (see e.g. Fig.~2 from
\cite{patat11}). Within this region there are also discrete O$_3$
bands, which can impact on the study of potential stellar features.
In the context of the potential diagnostic lines discussed here, it is
notable that one of the bands spans the (rest-frame) He~\2 \lam3203
line (e.g. Fig.~1 from \cite{m75}). Although weaker at the longer
wavelengths, careful correction for the O$_3$ bands will also be
required to study the O~\3 features discussed above,

On this topic we note the study by \cite{griffin05} who used historic
stellar spectroscopy to recover information on the intensity of past
O$_3$ features; this highlighted that with careful modelling (helped
in this particular case by near-UV HST observations) the stellar
features could be removed successfully to reveal the O$_3$ band
absorption; the reverse would also be true in terms of recovering the
intrinsic stellar spectra.

The operational concept for CUBES is under development as the project
progresses, but careful subtraction of the O$_3$ features will be an
important consideration. The current plans include observations of
flux standards (with high signal-to-noise) to help reconstruct the
atmospheric absorption features, combined with using theoretical tools
such as {\sc molecfit} \cite{molecfit}.

\section{Summary}

We have used model atmospheres to investigate the spectral diagnostics available over the 3000-4050\,\AA\ range that will be accessible with the new CUBES
instrument now in development for the VLT. There have been relatively
few studies of OB-type stars shortwards of the Balmer limit at
(ground) near UV wavelengths to date, but CUBES will provide an exciting new
capability to study massive stars in metal-poor systems at the edge of the 
Local Group and beyond.

Following the pioneering work from \cite{m75}, the He~\1 \lam3188 and
He~\2 \lam3203 pair of lines appear to be a compelling diagnostic of
stellar temperature, with the high Balmer series of lines providing
constraints on gravity. The near UV is also rich with metallic lines
(O~\3, O~\4, Si~\3, Si~\4, N~\3) that can provide further temperature
diagnostics (where more than one ionisation stage is present for a
given element) as well as estimates of chemical abundances in O-type
stars, particularly for oxygen where we lack robust diagnostics at
visible wavelengths.

Our results from the ETC presented here demonstrate that it should be
possible with CUBES to obtain high-quality (S/N\,$>$\,100,
$R$\,$\sim$\,7,000) near-UV spectra of massive O-type stars in
metal-poor systems at $\sim$1\,Mpc in $\sim$3\,hr integrations. The
CUBES analysis will probably be informed by initial estimates of
stellar parameters from observations at other wavelengths, and will
provide further constraints on stellar temperatures and gravities,
together with estimates of projected rotational velocities and the
first oxygen abundances for O-type stars in these extragalactic
systems. Such results will enable much needed comparisons with
theoretical predictions from low-metallicity evolutionary models and
perhaps the first observational evidence for chemically-homogeneous
evolution. Furthermore, using very metal-poor model spectra (with
$Z$\,$=$\,0.03$Z_{\odot}$) we have shown that CUBES could obtain
spectra with S/N\,$\sim$\,50 of the candidate O-type star in Leo~P at
1.6\,Mpc in approximately half a night of observations.

Our qualitative consideration of the models here was intended as a
first study of the potential diagnostics available in the CUBES
domain. Data from the ongoing XShootU ESO Large Programme (in support
of the {\em HST} ULLYSES Legacy Survey) will soon enable quantitative
investigation of O-type spectra in the CUBES region from high S/N
($>$\,100) X-Shooter observations in the Magellanic Clouds. Analysis
of the XShootU data will provide an important test of the temperatures
estimated using the He~\1~\lam3188/He~\2~\lam3203 ratio and log$g$
from the high Balmer lines compared to diagnostics in the visible. Our
expectation is that for the extragalactic metal-poor dwarfs envisaged
as future CUBES targets, the impact of stellar winds on the
determination of physical parameters from the near-UV region alone
should be relatively minor, and the XShootU data in the Clouds will
provide an important test of this.

More generally, the greater throughput and resolving power delivered
by CUBES compared to X-Shooter will give a unique capability for
studies of massive stars in the ground-UV region. Moreover, the
absence of cross-dispersion in the CUBES design will avoid the
challenges of blaze correction and order recombination that affect the
analysis of data from echelle instruments such as UVES and X-Shooter;
the CUBES design will give a smooth, continuous spectral response for
the two optical channels. This will enable more robust definition of the stellar
continuum for normalization, particularly for broad features in the
Balmer line series, as well as analysis of weaker absorption lines if
they would otherwise be in the reconnection regions. The latter will
be important for more rapidly-rotating massive stars
($v$sin$i$\,$\gtrsim$\,100\,\kms), which will also benefit from the
greater resolving power compared to X-Shooter.

\begin{acknowledgements}
We thank the reviewers for their suggestions on the manuscript, which
helped clarify details as well as helping place this study in the
wider context of previous work and some of the challenges of
ground-based observations at these short wavelengths. MG acknowledges
financial support from grants ESP2017-86582-C4-1-R and
PID2019-105552RB-C41, and from the Unidad de Excelencia “Marı\'{a} de
Maeztu” – Centro de Astrobiologı\'{a} (CSIC-INTA) project,
MDM-2017-0737.
\end{acknowledgements}


\section*{Conflict of interest}

The authors declare that they have no conflict of interest.

\section*{Data availability statement}
The {\sc tlusty} models used in this article are freely available online. The UVES data of HDE\,269896 and the {\sc cmfgen} models shown in Figs.~\ref{hd269896} and \ref{O_models} are available on request.


\begin{thebibliography}{}

\bibitem{wf00} Walborn, N. R. \& Fitzpatrick, E. L.: 
Contemporary Optical Spectral Classification of the OB Stars: A Digital Atlas.
PASP, 102, 379 (1990)

\bibitem{puls96} Puls, J., Kudritzki, R.-P., Herrero, A. et al.:
O-star mass-loss and wind momentum rates in the Galaxy and the Magellanic Clouds Observations and theoretical predictions.
A\&A, 305, 171 (1996)

\bibitem{zanutta22} Zanutta, A., Cristiani, S., Atkinson, D. et al.:
CUBES Phase A design overview.
ExA, 55, 241 (2023)

\bibitem{lamers72} Lamers, H. J.:
The spectrum of the supergiant $\epsilon$ Orionis (B0 Ia). I. Identifications, equivalent-widths, line profiles.
A\&AS, 7, 113 (1972)

\bibitem{m75} Morrison, N. D.:
The lines He~{\footnotesize I} $\lambda$3187 and He~{\footnotesize II} $\lambda$3203 in O-type stars
ApJ, 202, 433 (1975)
  
\bibitem{dufton} Dufton, P. L. \& McKeith, C. D.:
Copernicus observations of neutral helium lines in early-type stars.
A\&A, 81, 8 (1980)

\bibitem{drissen95} Drissen, L., Moffat, A. F. J., Walborn, N. R. \& Shara, M. M.:
The Dense Galactic Starburst NGC 3603. I. HST/FOS Spectroscopy of Individual Stars in the Core and the source of Ionization and Kinetic Energy.
AJ, 110, 2235 (1995)

\bibitem{wh00} Walborn, N. R. \& Howarth, I. D.:
Digital Spectroscopy of O3-O5 and ON/OC Supergiants in Cygnus.
PASP, 112, 1446 (2000)

\bibitem{ecfh04} Evans, C. J., Crowther, P. A., Fullerton, A. W. \& Hiller, D. J.:
Quantitative Studies of the Far-Ultraviolet, Ultraviolet, and Optical Spectra of Late O- and Early B-Type Supergiants in the Magellanic Clouds.
ApJ, 610, 1021 (2004)

\bibitem{cmfgen} Hillier, D. J. \& Miller, D. L.:
The Treatment of Non-LTE Line Blanketing in Spherically Expanding Outflows.
ApJ, 496, 407 (1998)

\bibitem{tlusty} Lanz, T. \& Hubeny, I.:
A Grid of Non-LTE Line-blanketed Model Atmospheres of O-Type Stars.
ApJS, 146, 417 (2003)

\bibitem{vfts} Evans, C. J., Lennon, D. J., Langer, N. et al.:
The VLT-FLAMES Tarantula Survey.
Msngr, 181, 22 (2020)

\bibitem{gs98} Grevesse, N. \& Sauval, A.:
Standard solar composition
SSRv, 85, 161, (1998)

\bibitem{cssj17} Sab{\'i}n-Sanjuli{\'a}n, C., Sim{\'o}n-D{\'i}az, S., Herrero, A. et al.:
The VLT-FLAMES Tarantula Survey. XXVI. Properties of the O-dwarf population in 30 Doradus.
A\&A, 601, A79 (2017)

\bibitem{trundle07} Trundle, C., Dufton, P. L., Hunter, I. et al.:
The VLT-FLAMES survey of massive stars: evolution of surface N abundances and effective temperature scales in the Galaxy and Magellanic Clouds.
A\&A, 471, 625 (2007)

\bibitem{sota11} Sota, A., Ma\'{i}z Apell\'{a}niz, J., Walborn, N. R. et al.:
The Galactic O-Star Spectroscopic Survey. I. Classification System and Bright Northern Stars in the Blue-violet at R$\sim$2500.
ApJS, 193, 24 (2011)

\bibitem{markova11} Markova, N., Puls, J., Scuderi, S. et al.:
Spectroscopic and physical parameters of Galactic O-type stars. I. Effects of rotation and spectral resolving power in the spectral classification of dwarfs and giants.
A\&A, 530, A11 (2011)

\bibitem{arias16} Arias, J. I., Walborn, N. R., Sim\'{o}n-D\'{i}az, S. et al.:
Spectral classification and properties of the O~Vz stars in the Galactic O-Star Spectroscopic Survey (GOSSS)
AJ, 152, 31 (2016)

\bibitem{ora17} Ram\'{i}rez-Agudelo, O. H., Sana, H., de Koter, A. et al.:
The VLT-FLAMES Tarantula Survey . XXIV. Stellar properties of the O-type giants and supergiants in 30 Doradus.
A\&A, 600, A81 (2017)

\bibitem{voyage2050} Garcia, M., Evans, C. J., Bestenlehner, J. M. et al.:
Massive stars in extremely metal-poor galaxies: a window into the past.
ExA, 51, 887 (2021)

\bibitem{IZw18} Sz\'{e}csi, D., Langer, N., Yoon, S.-C. et al.:
Low-metallicity massive single stars with rotation. Evolutionary models applicable to I Zwicky 18.
A\&A, 581, A15 (2015)

\bibitem{puls08} Puls, J., Vink, J. S. \& Najarro, F.:
Mass loss from hot massive stars.
A\&ARv, 16 209 (2008)

\bibitem{langer12} Langer, N.:
Presupernova Evolution of Massive Single and Binary Stars
ARA\&A, 50, 107, (2012)

\bibitem{mm01} Maeder, A. \& Meynet, G.:
Stellar evolution with rotation. VII. Low metallicity models and the blue to red supergiant ratio in the SMC.
A\&A, 373, 555 (2001)

\bibitem{brott11} Brott, I., de Mink, S. E., Cantiello, M. et al.:
Rotating massive main-sequence stars. I. Grids of evolutionary models and isochrones.
A\&A, 530, A115 (2011)

\bibitem{martins15} Martins, F., Herv\'{e}, A., Bouret, J.-C. et al.:
The MiMeS survey of magnetism in massive stars: CNO surface abundances of Galactic O stars.
A\&A, 575, A34 (2015)

\bibitem{mb22} Marcolino, W. L. F., Bouret, J.-C., Rocha-Pinto, H. J. et al.:
Wind properties of Milky Way and SMC massive stars: empirical Z dependence from CMFGEN models
MNRAS, 511, 5104 (2022)  

\bibitem{walborn2010} Walborn, N. R., Howarth, I. D., Evans, C. J. et al.:
The Onfp Class in the Magellanic Clouds.
AJ, 139, 1283 (2010)

\bibitem{bresolin06} Bresolin, F., Pietrzy\'{n}ski, G., Urbaneja, M. A. et al.:
The Araucaria Project: VLT Spectra of Blue Supergiants in WLM- Classification and First Abundances.
ApJ, 648, 1007 (2006) 
 
\bibitem{evans07} Evans, C. J., Bresolin, F., Urbaneja, M. A. et al.:
The ARAUCARIA Project: VLT-FORS Spectroscopy of Blue Supergiants in NGC 3109 -- Classifications, First Abundances, and Kinematics.
ApJ, 659, 1198 (2007)

\bibitem{castro55} Castro, N., Urbanejea, M. A., Herrero, A. et al.:
The ARAUCARIA project: Grid-based quantitative spectroscopic study of massive blue stars in NGC 55.
A\&A, 542, A79 (2012)

\bibitem{garcia13} Garcia, M. \& Herrero, A.:
The young stellar population of IC 1613. III. New O-type stars unveiled by GTC-OSIRIS.
A\&A, 551, A74 (2013)

\bibitem{Cal16} Camacho, I., Garcia, M., Herrero, A., \& Sim{\'o}n-D{\'{\i}}az, S.:
OB stars at the lowest Local Group metallicity. GTC-OSIRIS observations of Sextans A.
A\&A, 585, A82 (2016)

\bibitem{tramper11} Tramper, F., Sana, H., de Koter, A. \& Kaper, L.:
On the Mass-loss Rate of Massive Stars in the Low-metallicity Galaxies IC 1613, WLM, and NGC 3109.
ApJ, 741, L8 (2011)

\bibitem{tramper14} Tramper, F., Sana, H., de Koter, A. et al.:
The properties of ten O-type stars in the low-metallicity galaxies IC 1613, WLM, and NGC 3109.
A\&A, 572, A36 (2014)

\bibitem{rss13} Ram\'{i}rez-Agudelo, O. H., Sim\'{o}n-D\,{i}az, S., Sana, H. et al.:
The VLT-FLAMES Tarantula Survey. XII. Rotational velocities of the single O-type stars.
A\&A, 560, A29 (2013)

\bibitem{a09} Aerts, C., Puls, J., Godart, M. \& Dupret, M. A.:
Collective pulsational velocity broadening due to gravity modes as a physical explanation for macroturbulence in hot massive stars.
A\&A, 508, 409 (2009)

\bibitem{ssd10} Sim\'{o}n-D\'{i}az, S., Herrero, A., Uytterhoeven, K. et al.:
Observational Evidence for a Correlation Between Macroturbulent Broadening and Line-profile Variations in OB Supergiants.
ApJ, 720, L174 (2010)

\bibitem{ssd14} Sim\'{o}n-D\'{i}az, S. \& Herrero, A.:
The IACOB project. I. Rotational velocities in northern Galactic O- and early B-type stars revisited. The impact of other sources of line-broadening.
A\&A, 562, A135 (2014)

\bibitem{johnson} Johnson, H. L.:
Astronomical Measurements in the Infrared.
ARA\&A, 4, 193 (1966)

\bibitem{cardelli} Cardelli, J. A., Clayton, G. C. \& Mathis, J. S.:
The Relationship between Infrared, Optical, and Ultraviolet Extinction.
ApJ, 345, 245 (1989)

\bibitem{genoni22} Genoni, M., Landoni, M., Cupani, G. et al.:
The CUBES Instrument Model and Simulation Tools.
ExA, 55, 301 (2023)

\bibitem{leoP} Evans, C. J., Castro, N., Gonzalez, O. A. et al.:
First stellar spectroscopy in Leo P.
A\&A, 622, A129 (2019)

\bibitem{mcquinn} McQuinn, K. B. W., Skillman, E. D., Dolphin, A. et al.:
Leo P: An Unquenched Very Low-mass Galaxy.
ApJ, 812, 158 (2015)

\bibitem{skillman13} Skillman, E. D., Salzer, J. J., Berg, D. A. et al.:
ALFALFA Discovery of the nearby Gas-rich Dwarf Galaxy Leo P. III. An Extremely Metal Deficient Galaxy.
AJ, 146, 3 (2013)

\bibitem{telford21} Telford, G. O., Chisholm, J, McQuinn, K. B. W. \& Berg D.:
Far-ultraviolet Spectra of Main-sequence O Stars at Extremely Low Metallicity.
ApJ, 922, 191 (2021)

\bibitem{kud08} Kudritzki, R.-P., Urbaneja, M., Bresolin, F. et al.:
Quantitative Spectroscopy of 24 A Supergiants in the Sculptor Galaxy NGC 300: Flux-weighted Gravity-Luminosity Relationship, Metallicity, and Metallicity Gradient.
ApJ, 681, 269 (2008)

\bibitem{z09} Zorec, J., Cidale, L., Arias, M. L. et al.:
Fundamental parameters of B supergiants from the BCD system. I. Calibration of the (\lam$_1$, D) parameters into T$_{\rm eff}$.
A\&A, 501, 297 (2009)

\bibitem{s18} Shokry, A. Rivinius, Th, Mehner, A. et al.:
Stellar parameters of Be stars observed with X-shooter.
A\&A, 609, A108 (2018)

\bibitem{patat11} Patat, F., Moehler, S., O'Brien, K. et al.:
Optical atmospheric extinction over Cerro Paranal.
A\&A, 527, A91 (2011)

\bibitem{griffin05} Griffin, R. E.:
The detection and measurement of telluric ozone from stellar spectra.
PASP, 117, 885 (2005)
  
\bibitem{molecfit} Smette, A., Sana, H., Noll, S. et al.:
Molecfit: A general tool for telluric absorption correction. I. Method and application to ESO instruments.
A\&A, 576, A77 (2015)

\end{thebibliography}


\end{document}